\documentclass[12pt]{article}
\usepackage{amsmath,amssymb,amsfonts}

\begin{document}

\title{Electromagnetic waves and Stokes parameters in the wake of a gravitational wave}

\author{Shahen Hacyan
}

\renewcommand{\theequation}{\arabic{section}.\arabic{equation}}

\maketitle
\begin{center}

{\it  Instituto de F\'{\i}sica,} {\it Universidad Nacional Aut\'onoma de M\'exico,}

{\it Apdo. Postal 20-364, M\'exico D. F. 01000, Mexico.}


\end{center}
\vskip0.5cm

\begin{abstract}

A theoretical description of electromagnetic waves in the background of a (weak) gravitational wave is presented.
Explicit expressions are obtained for the Stokes parameters during the passage of a plane-fronted gravitational
wave described by the Ehlers-Kundt metric. In particular, it is shown that the axis of the polarization ellipse
oscillates, its ellipticity remaining constant.

\end{abstract}

PACS: 04.30.Nk; 42.25.Bs; 42.25.Ja


 \maketitle

\newpage
\section{Introduction}

In recent years there has been considerable interest in the detection of gravitational waves using interferometric
methods \cite{ligo}. Accordingly, it is convenient to have a precise description of how an electromagnetic (EM)
wave is affected by the passage of a gravitational wave. The goal of the present article is to describe this
interaction solving Maxwell's equations in the corresponding space-time background. For this purpose, the metric
of Ehlers and Kundt \cite{ehler} provides an appropriate description of a plane-fronted gravitational wave and
will be used throughout the paper. Particular attention will be focused on the Stokes parameters, as these are
directly observable and provide a complete description of the EM waves.

In section 2 of this article, Maxwell's equations in the background of the Ehlers-Kundt metric are solved for a
weak gravitational field; all relevant formulas for the EM field are then deduced within the short wave-length
approximation. Section 3 is devoted to the calculation of the Stokes parameters describing the polarization of the
EM wave. The main physical result is that the eccentricity of the polarization ellipse remains constant but its
axis rotates during the passage of a gravitational wave. If the latter has an oscillatory behavior, it will
produce an oscillation of the polarization ellipse axis with the same frequency.

\section{Plane-fronted gravitational waves}

The metric of a plane-fronted gravitational is given by \cite{ehler}
\begin{equation}
ds^2= g_{\mu \nu} dx^{\mu} dx^{\nu} = H(u,x,y) ~du^2 -2 du~dv + dx^2+dy^2~,
\end{equation}
where
\begin{equation}
H(u,x,y)= a(u) (x^2-y^2) +2b(u) xy~,
\end{equation}
and $a$ and $b$ are functions of the null-coordinate $u$. The inverse metric tensor is
\begin{equation}
g^{\mu \nu} =
\left( \begin{array}{cccc}
                    0 & -1 & 0 & 0 \\
                    -1 & -H & 0 & 0 \\
                    0 & 0 & 1 & 0 \\
                    0 & 0 & 0 & 1 \\
                  \end{array}
                \right).
\end{equation}
Here and in the following, coordinates are taken in the order $(u,v,x,y)$.

Minkowski space-time in Cartesian coordinates $(t, x,y,z)$ with signature $(-,+,+,+)$ is recovered if $H=0$. In
this limit, one can identify the coordinates as
$$
u= \frac{1}{\sqrt{2}}(t-z)~  ,\quad v= \frac{1}{\sqrt{2}}(t+z).
$$

As for the D'Alembertian operator, it takes the form
$$
\Box = \frac{\partial}{\partial x^{\mu}}\Big( g^{\mu \nu} \frac{\partial}{\partial x^{\mu}}\Big)=
$$
\begin{equation}
-2\frac{\partial^2}{\partial u~ \partial v} - H(u,x,y) \frac{\partial^2}{\partial v^2} +
\frac{\partial^2}{\partial x^2}+\frac{\partial^2}{\partial y^2}
\end{equation}
(notice that $\sqrt{-g}=1$).

The electromagnetic field tensor $F_{\alpha \beta}$ can be obtained from a potential $A_{\alpha}$ such that
\begin{equation}
F_{\alpha \beta}= \partial_{\alpha} A_{\beta} - \partial_{\beta} A_{\alpha},
\end{equation}
and must satisfy the Maxwell equations in vacuum,
\begin{equation}
\frac{\partial}{\partial x^{\beta}} F^{\alpha \beta} =0.
\end{equation}
A convenient choice for the gauge condition is $A_v=0$, since in this gauge the components of $A_{\mu}$ decouple
and the following explicit forms of the equations are obtained:
\begin{equation}
\Box A_x =0= \Box A_y~,
\end{equation}
\begin{equation}
\Box A_u = \partial_v (A_x\partial_x H  +A_y\partial_y H ),
\end{equation}
as can be checked by direct substitution. The last of these equations is equivalent to
\begin{equation}
 \partial_v A_u  -\partial_x A_x  -\partial_y A_y  =0,
\end{equation}
which is consistent with Maxwell's equations.

The general solution of the equation $\Box A=0$, valid to first order in the gravitational potentials, is of the
form
\begin{equation}
A \propto \exp \{ik_{\mu} x^{\mu} +i\Phi \},
\end{equation}
where
$$
\Phi (u,x,y) = - \frac{1}{2} k_v \Big[\alpha'' (x^2-y^2) + 2 \beta'' xy \Big]
$$
\begin{equation}
-(\alpha' k_x + \beta' k_y)~x + (-\beta' k_x +\alpha' k_y) ~y  - \frac{1}{k_v} \Big[\alpha (k_x^2 -k_y^2) + 2 \beta k_x k_y \Big],
\end{equation}
with
$$
a(u) = \alpha'''(u) , \quad b(u)= \beta'''(u)~,
$$
the primes denoting derivative with respect to $u$. Here and in the following, all quadratic or higher order terms
in $a$ and $b$ are to be dropped, in accordance with the weak field approximation. In the above equations,
$k_{\mu}$ can be identified with the four-vector of the EM field before the arrival of the gravitational wave
(that is, in flat space); its components are constant and satisfy the condition
\begin{equation}
-2 k_u~k_v + k_x^2+k_y^2=0.
\end{equation}

\subsection{Short wave-length approximation}

In order to proceed further, set
\begin{equation}
A_{\alpha} \equiv a_{\alpha} e^{i S},
\end{equation}
where $S$ is the eikonal function. Then Maxwell's equations $F_{\alpha \beta}^{~~;\beta}=0$ imply
$$
- \Big[a_{\alpha} S_{,\beta} - a_{\beta} S_{,\alpha}\Big]S^{,\beta}
$$
$$
+i \Big[a_{\alpha} S_{,\beta}^{~~;\beta} + 2 a_{\alpha ;\beta} S^{,\beta}-a_{\beta ;\alpha} S^{,\beta}-
a_{\beta}^{~;\beta} S_{,\alpha} -a^{\beta} S_{;\alpha \beta}\Big]
$$
\begin{equation}
+ a_{\alpha;\beta}^{~~~;\beta}-a_{\beta ;\alpha}^{~~~;\beta}=0.\label{om3}
\end{equation}

In the short wave-length approximation, the above equation is taken as an expansion in descending powers of $S$ (see,
e.g., Misner, Thorne and Wheeler \cite{mtw}, Sect. 22.5). Defining the null four-vector
$$K_{\alpha} = S_{,\alpha},$$
 the first term in the expansion implies
\begin{equation}
a^{\alpha} K_{\alpha} =0
\end{equation}
and the electromagnetic field turns out to be
\begin{equation}
F_{\alpha \beta} = i(K_{\alpha} a_{\beta} - K_{\beta}
a_{\alpha})e^{i S}.
\end{equation}

The second term in the expansion (\ref{om3}) implies
\begin{equation}
(a^2  K^{\alpha})_{,\alpha} =0,
\end{equation}
and  the energy-momentum tensor, to first order, is
\begin{equation}
T_{\alpha \beta} =  a^2  K_{\alpha} K_{\beta} . \label{Tnull}
\end{equation}

In the present case, the eikonal function is
\begin{equation}
S= k_{\mu} x^{\mu} + \Phi,\label{S}
\end{equation}
since
\begin{equation}
g^{\mu \nu} S_{,\mu} S_{,\nu} = 0,
\end{equation}
again to first order in $\alpha$ and $\beta$. Accordingly,
\begin{equation}
K_{\alpha} =
\begin{pmatrix}
  k_u + \partial_u \Phi\\
  k_v \\
  k_x - k_v (\alpha'' x +\beta'' y) -\alpha' k_x - \beta' k_y \\
  k_y - k_v (\beta'' x - \alpha'' y) -\beta' k_x +\alpha' k_y \label{K_a}
\end{pmatrix}
\end{equation}

As for the four-vector $a_{\mu}$, the components $a_x$ and $a_y$  must be independent of coordinates (though they
may depend in general on $k_{\mu}$), while $a_v=0$ and
\begin{equation}
a_u = k_v^{-1} (a_x K_x + a_y K_y).
\end{equation}

\section{Stokes parameters}

It is convenient to define time-like  and space-like unit four-vectors, $t_{\alpha}$ and $z_{\alpha}$
respectively, as
\begin{equation}
t_{\alpha} =\frac{1}{\sqrt{2}} \{-1+ \frac{1}{2}H,~ -1,~0,0\} \label{t}
\end{equation}
\begin{equation}
z_{\alpha} =\frac{1}{\sqrt{2}} \{-1 - \frac{1}{2}H,~ 1,~0,0\} \label{z},
\end{equation}
such that $t^{\alpha} t_{\alpha}=-1$, $z^{\alpha} z_{\alpha}=1$ and $t^{\alpha} z_{\alpha}=0$. Then the frequency
$\Omega$ of the EM wave as measured by an observer at rest with unit velocity four-vector $t_{\alpha}$ is
\begin{equation}
\Omega \equiv  -K_{\alpha} t^{\alpha} = -\frac{1}{\sqrt{2}} [K_u + (1+ \frac{1}{2}H) k_v].
\end{equation}
Clearly $\Omega$ depends on position and time. It is also convenient to define
\begin{equation}
K_{\|} = K_{\alpha} z^{\alpha} ,
\end{equation}
and then $\Omega^2 = K_{\|}^2 + K_{\bot}^2$, where $K_{\bot}= \sqrt{K_x^2 + K_y^2}$.

Notice that in flat space, the frequency of the EM wave is just $$\omega =-\frac{1}{\sqrt{2}} (k_u + k_v) $$ and
the $z$ component of the wave-vector is
$$k_z = \frac{1}{\sqrt{2}}(-k_u +k_v).$$

We can now define two four-vectors $\hat{\epsilon}_{\alpha}^{(i)}$ ($i=1,2$) orthogonal to $t^{\alpha}$ and
$K^{\alpha}$:
\begin{eqnarray}
\hat{\epsilon}_{\alpha}^{(1)} &=& K_{\bot}^{-1} \{ 0,~0,- K_y,~ K_x  \} \nonumber \\
\hat{\epsilon}^{(2)\alpha} &=& - \frac{K_{\bot}}{\Omega} z_{\alpha} + \frac{K_{\|}}{\Omega K_{\bot}} \{0,~0,~K_x,
K_y\}. \\ \nonumber
\end{eqnarray}
The electric field in the frame at rest is defined quite generally as
\begin{equation}
E^{\alpha} = F^{\alpha \beta} t_{\beta},
\end{equation}
and accordingly the Stokes parameters can be constructed from the two scalar products
$E^{\alpha}\hat{\epsilon}_{\alpha}^{(i)}$. Namely
\begin{eqnarray}
S_0 &=& |E^{\alpha}\hat{\epsilon}_{\alpha}^{(1)}|^2 + |E^{\alpha}\hat{\epsilon}_{\alpha}^{(2)}|^2 \nonumber \\
S_1 &=& |E^{\alpha}\hat{\epsilon}_{\alpha}^{(1)}|^2 - |E^{\alpha}\hat{\epsilon}_{\alpha}^{(2)}|^2 \nonumber \\
S_2 +i S_3&=& 2 (E^{\alpha}\hat{\epsilon}_{\alpha}^{(1)})^* E^{\alpha}\hat{\epsilon}_{\alpha}^{(2)} ,
\end{eqnarray}
following the notation of Born and Wolf \cite{BW}.

Now,
\begin{equation}
E_{\alpha} = i [\Omega a_{\alpha} + (a_{\beta} t^{\beta}) K_{\alpha}]e^{iS}
\end{equation}
and since
$$
K_{\alpha} \hat{\epsilon}^{(i)\alpha} =0,
$$
it follows that
\begin{equation}
E_{\alpha} \hat{\epsilon}^{(i)\alpha} = i \Omega a_{\alpha} \hat{\epsilon}^{(i)\alpha} e^{iS},
\end{equation}
or explicitly
\begin{equation}
E_{\alpha} \hat{\epsilon}^{(1)\alpha} = i \frac{\Omega}{K_{\bot}}(a_y K_x  -a_x K_y ) e^{iS}~,
\end{equation}
\begin{equation}
E_{\alpha} \hat{\epsilon}^{(2)\alpha} = -i \frac{\Omega}{K_{\bot}}(a_x K_x  +a_y K_y ) e^{iS}.
\end{equation}

For algebraic calculations, it is convenient to express the Stokes parameters in matrix form as
\begin{equation}
\left(
  \begin{array}{cc}
    S_0 + S_1 & S_2 -i S_3 \\
    S_2 +i S_3  & S_0 - S_1 \\
  \end{array}
\right) = 2 \frac{\Omega^2}{K_{\bot}^2}  \mathbb{K}^T \left(
  \begin{array}{cc}
    |a_x|^2 & a_x a_y^* \\
    a_x^* a_y  & |a_y|^2 \\
  \end{array}
\right)
\mathbb{K},
\end{equation}
where
\begin{equation}
\mathbb{K}=
\left(
  \begin{array}{cc}
    K_y & K_x \\
    -K_x & K_y \\
  \end{array}
\right).
\end{equation}
The component $K_x$ and $K_y$ are given by (\ref{K_a}).

The important point is that $a_x$ and $a_y$ are constant, and therefore these quantities can be expressed in terms
of their values before the arrival of the gravitational wave or, equivalently, in terms of the Stokes parameters
$s_i$ in flat space. The following relations are obtained with some straightforward algebra:
\begin{eqnarray}
S_0 &=& \frac{\Omega^2}{\omega^2} s_0 \nonumber \\
S_1 &=& \frac{\Omega^2}{\omega^2}  (s_1 \cos 2\delta - s_2 \sin 2\delta ) \nonumber \\
S_2 &=& \frac{\Omega^2}{\omega^2} (s_1 \sin 2\delta + s_2 \cos 2\delta ) \nonumber \\
S_3 &=& \frac{\Omega^2}{\omega^2} s_3 ~,
\end{eqnarray}
where
\begin{equation}
e^{i \delta} = \frac{1}{k_{\bot}K_{\bot}} (k_x -ik_y)(K_x +i K_y ).
\end{equation}

The Stokes parameters can also be expressed in terms of two angles, $\chi$ and $\psi$,  spanning the
Poincar\'e sphere (Born and Wolf \cite{BW}):
\begin{eqnarray}
S_1 &=& S_0 \cos 2\chi ~\cos 2\psi \nonumber \\
S_2 &=& S_0 \cos 2\chi ~\sin 2\psi \nonumber \\
S_3 &=& S_0 \sin 2\chi.
\end{eqnarray}
The angle $\psi$ specifies the orientation of the polarization ellipse and the angle $\chi$ its ellipticity.

The intensity of the EM wave before and during the passage of the gravitational wave is given by the parameters
$s_0$ and $S_0$ respectively. From the definition of $\Omega$, it follows that
\begin{equation}
\Omega -\omega = \frac{1}{\omega +k_z}\Big\{k_v [(k_x \alpha'' +k_y \beta'')x +(k_x \beta'' - k_y \alpha'')y ]+
\alpha' (k_x^2 -k_y^2) +2\beta' k_x k_y\Big\}.\label{38}
\end{equation}
Similarly, it can be seen that the angle $\chi$ remains constant, but the angle $\psi$ changes from $\psi_0$ to
$\psi_0 + \delta$ in the wake of the gravitational wave, where
\begin{equation}
\delta = - \frac{1}{k_{\bot}^2}\Big\{k_v [(k_y \alpha'' -k_x \beta'')x +(k_x \alpha'' + k_y \beta'')y ] - \beta'
(k_x^2 -k_y^2) +2\alpha' k_x k_y\Big\}.\label{39}
\end{equation}

It follows from the above formulas that the frequency of the EM wave is shifted from $\omega$ to $\Omega$. As for
the polarization, a circularly polarized wave ($s_1=s_2=0$) remains circularly polarized, but an elliptically or
linearly polarized wave changes its direction of polarization. The above formulas show that the ellipticity of the
EM wave is not altered, but the angle $\psi$, defining the axis of the ellipse, oscillates with the frequency of
the functions $a(u)$ and $b(u)$.

\section{Summary of results}

We have obtained the relevant equations describing an EM wave in the presence of a plane-fronted gravitational
wave in terms of the eikonal function $S$ given by Eq. (\ref{S}) and two functions, $\alpha (u)$ and $\beta (u)$,
characterizing the gravitational wave .

The main result of the present article is that a gravitational wave interacting with a monochromatic EM wave
rotates its axis of polarization by an angle given by Eq. (\ref{39}), where the coefficients $k_{\mu}$ are the
components of the four-vector defining the EM wave before the arrival of the gravitational wave; this wave
four-vector changes to the form given by Eq. (\ref{K_a}) in the presence of the latter. The rotation is
oscillatory with the same frequency as that of the gravitational wave.

\section*{Acknowledgment}

Work supported in part by PAPIIT Project IN-101511 (DGAPA, UNAM).

\subsection*{Appendix}

In this appendix, the EM field in the limit of flat space-time is worked out according to the formalism of the
present paper. This permits to characterize the EM field before and during the passage of the gravitational wave
and compare both cases. In flat space-time, the electric field is given by
\begin{equation}
{\bf E} =i(\omega {\bf a} - a_z {\bf k})e^{iS},
\end{equation}
where $k^{\alpha} = ( \omega ,{\bf k})$ in Minkowski coordinates and $S =k_{\alpha} x^{\alpha}$. It is important
to notice that the gauge used in this paper is such that $a_v =-a^u=0$, and therefore $a^t = a^z$. Thus the
condition $k^{\alpha} a_{\alpha}=0$ implies
\begin{equation}
k_x a_x + k_y a_y = (\omega-k_z) a_z,
\end{equation}
and of course ${\bf E} \cdot {\bf k} =0$.

The unit vectors $\hat{\epsilon}_{\alpha}^{(i)}$ have purely space components:
\begin{eqnarray}
\hat{ \epsilon   }^{(1)} &=& k_{\bot}^{-1} (-k_y,k_x,0) \nonumber \\
\hat{ \epsilon   }^{(2)} &=& \omega^{-1} \Big(\frac{k_z}{k_{\bot}} k_x, \frac{k_z}{k_{\bot}} k_y, -k_{\bot}\Big)
\end{eqnarray}
and thus
\begin{eqnarray}
\hat{ \epsilon   }^{(1)} \cdot {\bf E} &=& i\omega k_{\bot}^{-1} (k_x a_y-k_y a_x ) e^{iS} \nonumber \\
\hat{ \epsilon   }^{(2)} \cdot {\bf E} &=&  -i\omega k_{\bot}^{-1} (k_x a_x +k_y a_y) e^{iS}.
\end{eqnarray}

\end{document}